\documentclass[traditabstract]{aa}

\usepackage{amsmath}
\usepackage{amssymb}
\usepackage{txfonts}
\usepackage{epsfig,graphicx}
\usepackage[colorlinks=true, linkcolor=blue, citecolor=blue, urlcolor=blue]{hyperref}
\usepackage{natbib}


\bibpunct{(}{)}{;}{a}{}{,} 

\begin{document}

\title{Analytical properties of Einasto dark matter haloes}
\titlerunning{Analytical properties of Einasto dark matter haloes}

\author{E. Retana-Montenegro \inst{1} 
\and E. Van Hese \inst{2}
\and G. Gentile \inst{2} 
\and M. Baes \inst{2}
\and F. Frutos-Alfaro \inst{1}} 

\institute{Escuela de F\'{i}sica, Universidad de Costa Rica, San Pedro 11501, Costa Rica\\
\email{edwin@fisica.ucr.ac.cr}
\and Sterrenkundig Observatorium, Universiteit Gent, Krijgslaan 281-S9, B-9000 Gent, Belgium\\
\email{maarten.baes@ugent.be}
}

\date{Received ...; Accepted...}

\date{}

\abstract{Recent high-resolution $N$-body CDM simulations indicate
  that nonsingular three-parameter models such as the Einasto profile
  perform better than the singular two-parameter models, e.g.~the
  Navarro, Frenk and White, in fitting a wide range of dark matter
  haloes.  While many of the basic properties of the Einasto profile
  have been discussed in previous studies, a number of analytical
  properties are still not investigated. In particular, a general
  analytical formula for the surface density, an important quantity
  that defines the lensing properties of a dark matter halo, is still
  lacking to date. To this aim, we used a Mellin integral transform
  formalism to derive a closed expression for the Einasto surface
  density and related properties in terms of the Fox $H$ and Meijer
  $G$ functions, which can be written as series expansions. This
  enables arbitrary-precision calculations of the surface density and
  the lensing properties of realistic dark matter halo
  models. Furthermore, we compared the S\'ersic and Einasto surface
  mass densities and found differences between them, which implies
  that the lensing properties for both profiles differ.}

\keywords{methods: analytical, gravitational lensing: strong, gravitational lensing: weak, galaxies: clusters: general, galaxies: halos, cosmology: dark matter.}

\maketitle

\section{Introduction}

The $\Lambda$ cold dark matter (CDM) model, with
$\left(\Omega_{m},\,\Omega_{\Lambda}\right)=\left(0.3,\,0.7\right)$,
has become the standard theory of cosmological structure formation.
$\Lambda$CDM seems to agree with the observations on
cluster-sized scales \citep{2003NuPhS.124....3P}; however, on
galaxy/sub-galaxy scales there appears to be a discrepancy between
observations and numerical simulations.  High-resolution observations
of rotation curves, in particular of low surface brightness (LSB) and
dark matter dominated dwarf galaxies
\citep{2001ApJ...552L..23D,2001MNRAS.325.1017V,2003ApJ...583..732S,2003MNRAS.340...12W,2004MNRAS.353L..17D,2005ApJ...634L.145G,2005ApJ...621..757S,2007MNRAS.375..199G,2010NewA...15...89B}
favour density profiles with a flat central core
(e.g. \citealt{1995ApJ...447L..25B,2000ApJ...537L...9S,2004MNRAS.351..903G,2009RAA.....9.1173L}).
In contrast, $N$-body (dark matter only) CDM simulations predict
galactic density profiles that are too high in the centre
(e.g. \citet*{1996ApJ...462..563N,1997ApJ...490..493N} (NFW);
\citealt{1999MNRAS.310.1147M}). This discrepancy is called the
cusp-core problem; a complete review can be found in
\citet{2010AdAst2010E...5D}.

Gravitational lensing is one of the most powerful tools in
observational cosmology for probing the matter distribution of
galaxies and clusters in the strong regime
\citep{1989MNRAS.238...43K,1994AJ....108.1156W,1996A&A...313..697B,1998ApJ...495..609C,2000ApJ...535..692K,2001ApJ...549L..25K,2002ApJ...574L.129S,2002ApJ...575L...1K,2003ApJ...582...17K,2004ApJ...612..660K,2008A&A...489...23L,2009A&A...507...35A,2011MNRAS.413.1753Z,2011arXiv1107.2649Z}
and the weak regime
\citep{1993ApJ...404..441K,1999ARA&A..37..127M,2001PhR...340..291B,2004ApJ...606...67H,2006ApJ...648L.109C,2007ApJ...668..806M,2009ApJ...704..672J,2011A&A...529A..93H}.
Comparing these observations to theoretical models provides key
information to help resolve the cusp-core problem. Evidently, one must
use the most accurate density profile to obtain the best fit to
observational data from strong- and weak-lensing studies.

Recently, $N$-body CDM simulations
\citep{2004MNRAS.349.1039N,2006AJ....132.2685M,2008MNRAS.387..536G,2008MNRAS.388....2H,2009MNRAS.398L..21S,2010MNRAS.402...21N}
have found that certain three-parameter profiles provide an excellent
fit to a wide range of dark matter haloes. One of these is the
\citet{1965TIAAA.17..01} profile, a three-dimensional version of the
two-dimensional \citet{1968adga.book.....S} profile used to describe
the surface brightness of early-type galaxies and the bulges of spiral
galaxies
\citep[e.g.][]{1988MNRAS.232..239D,1993MNRAS.265.1013C,1994MNRAS.271..523D,1994ApJS...93..397C,1995MNRAS.275..874A,1997A&A...321..111P,2001A&A...368...16M,2003AJ....125.2936G,2006AJ....132.2701G,2009MNRAS.393.1531G}.
The S\'ersic profile can be written as:
\begin{equation}
  \Sigma_{\text{S}}\left(R\right) = 
  \Upsilon I_{{\text{e}}}\exp\left\{ -b_{m}\left[\left(\frac{R}{R_{{\text{e}}}}\right)^{1/m}-
      1\right]\,\right\} ,\label{eq:Sersic-surface-mass-density}
\end{equation}
where $R$ is the distance in the sky plane, $m$ the S\'ersic index,
$\Upsilon$ is the mass-to-light ratio, {$I_{{\text{e}}}$} is the
luminosity density at the effective radius $R_{{\text{e}}}$, and
$b_{m}$ is a dimensionless function of $m$ that can be determined from
the condition that the luminosity inside $R_{{\text{e}}}$ equals half
of the total luminosity. Numerical solutions for $b_{m}$ are given by
\citet{1991A&A...249...99C}, \citet{1998A&AS..130...81M},
\citet{1997A&A...321..111P} and an asymptotic expansion
$b_{m}=2m-1/3+4/405m+\mathcal{O}\left(m^{2}\right)$ using analytical
methods was obtained by \citet{1999A&A...352..447C}.

The Einasto profile (model) is characterised by a power-law logarithmic
slope,
\begin{equation}
  \gamma(r)\equiv-\frac{{\text{d}}\ln\rho}{{\text{d}}\ln r}(r)\propto r^{1/n},
\end{equation}
with $n$, which we call the Einasto index, a positive number
defining the steepness of the power law. Integrating leads to the
general density profile
\begin{equation}
  \rho(r)=\rho_{\text{s}}\exp\left\{ -d_{n}\left[\left(\frac{r}{{r_{{\text{s}}}}}\right)^{1/n}-1\right]\,\right\},
\end{equation}
where ${r_{{\text{s}}}}$ represents the radius of
the sphere that contains half of the total mass, $\rho_{\text{s}}$
is the mass density at $r={r_{{\text{s}}}}$, and $d_{n}$
is a numerical constant that ensures that ${r_{{\text{s}}}}$
is indeed the half-mass radius. In the context of dark matter haloes,
the density can also be expressed as
\begin{equation}
  \rho\left(r\right) = 
  \rho_{-2}\,\exp\left\{ -2n\left[\left(\frac{r}{r_{-2}}\right)^{1/n}-1\right]\,\right\},
  \label{eq:einasto_halo_vers}
\end{equation}
where $\rho_{-2}$ and $r_{-2}$ are the density and radius
at which $\rho\left(r\right)\propto r^{-2}$. In the remainder of this
paper, we will use yet another, equivalent version, 
\begin{equation}
  \rho(r)=\rho_{0}\exp\left[-\left(\frac{r}{h}\right)^{1/n}\right],\label{eq:einasto_gen_vers}
\end{equation}
where we introduced the central density 
\begin{equation}
\rho_{0}=\rho_{s}\, e^{d_{n}}=\rho_{-2}\, e^{2n},
\end{equation}
and the scale length 
\begin{equation}
h=\frac{{r_{{\text{s}}}}}{d_{n}^{n}}=\frac{r_{-2}}{(2n)^{n}}.
\end{equation}
If a model is to describe real galactic systems, several conditions
must be imposed on the model description functions, as discussed by
\citet{1969AN....291...97E}. When a model is constructed, an initial
descriptive function is chosen; the most practical choice is the
density profile, because the main descriptive functions (cumulative
mass profile, gravitational potential, surface mass density) are
integrals of the density profile. Furthermore, a physical model has to
satisfy several conditions: i) the density profile must be
non-negative and finite; ii) the density must be a smoothly decreasing
function that approaches zero at large radii; iii) some moments of the
mass function must be finite, in particular moments that define the
central gravitational potential, the total mass, and the effective
radius of the system; and iv) the descriptive functions must not have
jump discontinuities.  \citet{1969AN....291...97E} presented several
families of valid descriptive functions, among which the so-called
Einasto profile is a special case that agrees best with observations.

The Einasto profile was used by \citet{1969Ap......5...67E}
to obtain a model of M31. Later, this model was applied by \citet{1974smws.conf..291E}
to several nearby galaxies, including M32, M87, Fornax and
Sculptor dwarfs, and also M31 and the Milky Way. These models were
multi-component ones, and each component has its own parameter set $\left\{ \,\rho_{0},\, h,\, n\,\right\} $
representing certain physically homogeneous stellar populations.

\citet{2004MNRAS.349.1039N} found that for haloes with masses from
dwarfs to clusters, $4.54\lesssim n\lesssim8.33$ with an average value
of $n=5.88$. \citet{2008MNRAS.388....2H} and
\citet{2008MNRAS.387..536G} found that $n$ tends to decrease with mass
and redshift, with $n\sim5.88$ for galaxy-sized and $n\sim4.35$ for
cluster-sized haloes in the Millennium Run (MR)
\citep{2005Natur.435..629S}. \citet{2010MNRAS.402...21N} obtained
similar results for galaxy-sized haloes in the Aquarius simulation
\citep{2008MNRAS.391.1685S}. Also, \citet{2008MNRAS.387..536G} showed
that $n\sim3.33$ for the most massive haloes of
MS. \citet{2011AJ....142..109C} modelled the rotation curves of a
spiral galaxies subsample from THINGS (The {\small HI} Nearby Galaxy
Survey, \citep{2008AJ....136.2648D}), using the Einasto profile, and
found that $n$ tends to have lower values by a factor of two or more
compared to those predicted by $N$-body
simulations. \citet{2011arXiv1112.3120D} fitted the surface brightness
profiles of a large sample of elliptical galaxies in the Virgo
cluster, using a multi-component Einasto profile consisting of two or
three components for each galaxy with different $n$ for each
individual component. For the central component, they found values of
$n\lesssim1$ for the most massive and shallow-cusp galaxies and $n<2$
for steep-cusp galaxies, while $5\lesssim n\lesssim8$ for the outer
component of massive galaxies. These values for the outer component
are consistent with the results from $N$-body simulations for Einasto
dark matter haloes, and the authors argued that Einasto components
with an $n$ in this range could be dark matter dominated.

In the light of its increasing popularity to describe the density
of simulated dark matter haloes, a detailed investigation of the properties
of the Einasto model is of paramount importance. Some aspects of the Einasto
model have been presented by several authors \citep{2005MNRAS.362...95M,2005MNRAS.358.1325C,2006AJ....132.2685M,2010MNRAS.405..340D}.
The most complete study of the properties of the Einasto model is
the work by \citet{2005MNRAS.358.1325C}, who provided a set of analytical
expressions for quantities such as the mass profile and gravitational
potential and discussed the dynamical structure for both isotropic and
anisotropic cases. Nevertheless, the Einasto model has not been studied
analytically as extensively as the S\'ersic models, and several properties
still have to be further investigated in more detail. One area where
progress is still to be made is on the value of the dimensionless
parameter $d_{n}$.

The most important lacuna, however, concerns the surface density on
the plane of the sky, an important quantity that defines the lensing
properties of a dark matter halo. The surface mass density has an
importance in theoretical predictions and observations. On the
observational side, based mostly on the mass decomposition of rotation
curves using cored haloes, various studies
\citep{2004IAUS..220..377K,2008MNRAS.383..297S,2009MNRAS.397.1169D,2009Natur.461..627G}
have shown that the product of the central density $\rho_{0}$ and the
core radius $r_{0}$ is consistent with a universal value, independent
of galaxy mass. The product $\rho_{0}r_{0}$ is directly proportional
to the average surface density within $r_{0}$, and to the
gravitational acceleration of dark matter at $r_{0}$. In view of a
detailed comparison of these results with the outcome of the most
recent dark matter simulations, it is therefore important to study the
analytical properties of the surface density distribution of Einasto
haloes. Recent works studying theoretical and observational aspects of
dark matter surface densities include \citet{2010PhRvL.104s1301B},
\citet{2010ApJ...717L..87W}, \citet{2010MNRAS.405.2351N}, and
\citet{2011MNRAS.tmp.1234C}: efforts are being made to confirm or call
into question the universality of the dark matter surface density
within one dark halo scale length.

\citet{2005MNRAS.358.1325C} showed that the surface density of the
Einasto model cannot be expressed in terms of elementary functions and
discussed the general properties using numerical
integration. \citet{2010MNRAS.405..340D} presented an analytical
approximation for the Einasto surface brightness profile and
demonstrated that this surface density profile is not S\'ersic-like.
A general analytical formula, which would enable an
arbitrary-precision calculation and an analytical study of the
asymptotic behaviour, is still lacking to date.

The most recent studies in strong and weak lensing
(e.g. \citealt{2011A&A...528A..73D,2011MNRAS.tmp.1281S,2011MNRAS.tmp.1336M,2011ApJ...738...41U,2011arXiv1109.2594O})
use the NFW profile or its generalization
\citep{1996MNRAS.278..488Z,2000ApJ...529L..69J} to model the dark
matter halo instead of the Einasto profile. One of the factors that
limits its adoption is the absence of analytical formulas for its
lensing properties. Therefore, its application in lensing studies is
not as wide as for other profiles. A complete general set of
analytical formulas for the lensing properties of the Einasto profile
would help to increase its use in lensing studies.

In this paper, we extend the analytical study of the Einasto model
using some of the techniques also employed in the extensive literature
on the analytical properties of the S\'ersic model (e.g.\
\citealt{1991A&A...249...99C,1997A&A...321..724C,1999A&A...352..447C,2001MNRAS.326..869T,2002A&A...383..384M,2004A&A...415..839C,2005PASA...22..118G,2007JCAP...07..006E,2011A&A...525A.136B,2011A&A...534A..69B}).
In Section \ref{sec:02}, we discuss an analytical expansion for the
dimensionless parameter $d_{n}$. In Section \ref{sec:03}, we derive an
analytical expression for the Einasto surface mass density in terms of
the Fox $H$ function, using the Mellin transform-method. We then use
the result for the projected surface mass density to calculate the
cumulative surface mass, lens equation, deflection angle, deflection
potential, magnification, shear and the critical curves for a
spherically symmetric lens described by the Einasto profile in terms
of this function.  We also calculate explicit series expansions for
the surface mass density and all the lensing properties. In Section
\ref{sec:04}, we derive some special cases for the lensing properties
for integer and half-integer values of $n$ in terms of the Meijer $G$
function.  Furthermore, we compare the Einasto and S\'ersic surface
mass densities for the same values of their respective
indices. Finally, in Section \ref{sec:05} we discuss the asymptotic
behaviour of the surface mass density and cumulative surface mass
at small and large radii. We present our conclusions in Section
\ref{sec:06}.

\section{Spatial properties\label{sec:02} }

The total mass of an Einasto model with a mass density profile given
by (\ref{eq:einasto_gen_vers}), is
\begin{equation}
  M=4\pi\,\rho_{0}\, h^{3}\, n\,\Gamma(3n).\label{Mtot}
\end{equation}
If we use this formula to replace the central density $\rho_{0}$
by the total mass $M$ as a parameter in the definition of the Einasto
models, we obtain 
\begin{equation}
\rho(r)=\frac{M}{4\pi\, h^{3}\, n\,\Gamma(3n)}\,{\text{e}}^{-s^{1/n}},
\end{equation}
where we have introduced the reduced radius 
\begin{equation}
  s=\frac{\left(d_{n}\right)^{n}\, r}{r_{s}}.
\end{equation}
At small radii, the density profile behaves as 
\begin{equation}
\rho(r)=\frac{M}{4\pi\, h^{3}\, n\,\Gamma(3n)}\left(1-s^{1/n}+\cdots\right).
\end{equation}
The cumulative mass profile $M(r)$ of a spherical mass distribution
is found through the equation 
\begin{equation}
M(r)=4\pi\int_{0}^{r}\rho(r')\, r'^{2}\,{\text{d}}r'.
\end{equation}
For the Einasto model, we find 
\begin{equation}
M(r)=M\left[1-\frac{\Gamma(3n,s^{1/n})}{\Gamma(3n)}\right],
\end{equation}
where $\Gamma(\alpha,x)$ is the incomplete gamma function, 
\begin{equation}
\Gamma(\alpha,x)=\int_{x}^{\infty}t^{\alpha-1}\,{\text{e}}^{-t}\,{\text{d}}t.
\end{equation}
Given that this is the radius of the sphere that encloses half of the
total mass, we find that $d_{n}$ is the solution of the equation
\begin{equation}
2\,\Gamma(3n,d_{n})=\Gamma(3n).
\end{equation}
This equation cannot be solved in a closed form; one option is to
solve it numerically or use interpolation formulae. For example,
\citet{2006AJ....132.2685M}, quoting G. A. Mamon (private
communication), propose $d_{n}\approx3n-\frac{1}{3}+0.0079/n$, but do
not mention the origin of this approximation or quote its
accuracy. An elegant way to determine an approximation for $d_{n}$
builds on the work by \citet{1999A&A...352..447C}, who used an
asymptotic expansion method to solve the general relation
$\Gamma(\alpha,x)=\Gamma(\alpha)/2$.  If we apply the resulting
expansion to our problem, we obtain for the parameter $d_{n}$ from the
Einasto model
\begin{multline}
  d_{n} \approx 3n-\frac{1}{3}+\frac{8}{1215\, n}+\frac{184}{229635\, n^{2}}\\
  +\frac{1048}{31000725\, n^{3}}-\frac{17557576}{1242974068875\, n^{4}}+{\cal \mathcal{O}}\left(\frac{1}{n^{5}}\right).\label{dn}
\end{multline}
The gravitational potential of a spherically symmetric mass distribution
$\rho(r)$ can be found through the formula 
\begin{equation}
  \Psi(r) = 4\pi G\left[\frac{1}{r}\int_{0}^{r}\rho(r')\, r'^{2}\,{\text{d}}r'+\int_{r}^{\infty}\rho(r')\, r'\,{\text{d}}r'\right],
\end{equation}
or equivalently 
\begin{equation}
\Psi(r)=\int_{r}^{\infty}\frac{M(r')\,{\text{d}}r'}{r'^{2}}.
\end{equation}
For the Einasto model, we find 
\begin{equation}
  \Psi(r) = \frac{GM}{h}\, s^{-1}\left[1-\frac{\Gamma(3n,s^{1/n})}{\Gamma(3n)}+\frac{s\,\Gamma(2n,s^{1/n})}{\Gamma(3n)}\right].
\end{equation}
The Einasto model has a finite potential well, given by 
\begin{equation}
  \Psi_{0} = \frac{GM}{h}\,\frac{\Gamma(2n)}{\Gamma(3n)}.
\end{equation}
At small radii, the potential decreases parabolically 
\begin{equation}
  \Psi(r) \sim \frac{GM}{h}\,\left[\frac{\Gamma(2n)}{\Gamma(3n)}-\frac{s^{2}}{6n\,\Gamma(3n)}+\cdots\right],
\end{equation}
which is not surprising, given the finite density core of the Einasto
density profile. At large radii, we obtain a Keplerian fall-off, 
\begin{equation}
  \Psi(r) \sim \frac{GM}{r}+\cdots,
\end{equation}
which is also expected for a model with a finite total mass.

\section{Lensing properties\label{sec:03}}

\subsection{Surface mass density and cumulative surface mass density}

The surface mass density of a spherically symmetric lens is given
by integrating along the line of sight of the 3D density profile
\begin{equation}
\Sigma\left(\xi\right) = \int_{-\infty}^{+\infty}\rho\left(\xi,r\right)\,{\text{d}}z,\label{eq:03}
\end{equation}
where $\xi$ is the radius measured from the centre of the lens and
$r=\sqrt{\xi^{2}+z^{2}}$. This expression can also be written as
an Abel transform \citep{1987gady.book.....B}
\begin{equation}
\Sigma\left(\xi\right)=2\,\int_{\xi}^{\infty}\frac{\rho\left(r\right)\, r\,{\text{d}}r}{\sqrt{r^{2}-\xi^{2}}}.\label{eq:04}
\end{equation}
Inserting eq.~(\ref{eq:einasto_gen_vers}) into the above expression,
we obtain
\begin{equation}
\Sigma\left(x\right) = 2\,\rho_{0}\, h\,\int_{x}^{\infty}\frac{{\text{e}}^{-s^{1/n}}\, s\,{\text{d}}s}{\sqrt{s^{2}-x^{2}}},\label{eq:Sigma_integral}
\end{equation}
where we have introduced the quantities $x=\xi/h$ and $s=r/h$.

As discussed by \citet{2005MNRAS.358.1325C} and
\citet{2010MNRAS.405..340D}, the integral (\ref{eq:Sigma_integral})
cannot be expressed in terms of elementary or even the most regular
functions for all the values of $n$. Only the central surface mass
density can be evaluated analytically as
\begin{equation}
  \Sigma(0) = 2n\,\rho_{0}\, h\,\Gamma\left(n\right).\label{Sigma0}
\end{equation}
This situation is very similar to the deprojection of the S\'ersic
surface brightness profile (\ref{eq:Sersic-surface-mass-density}).
Indeed, the deprojection formula leads to an integral that is an inverse
Abel transform, and also in this case it was long thought that no further
analytical progress was possible. However, \citet{2002A&A...383..384M}
used the computer package \texttt{{Mathematica}} to demonstrate
that it is possible to write the S\'ersic luminosity density in terms
of the Meijer $G$ function for all integer S\'ersic indices $m$.
\citet{2011A&A...525A.136B} and \citet{2011A&A...534A..69B} took
this analysis one step further and showed that the deprojection of
the S\'ersic surface brightness profile for general values of $m$
can be solved elegantly using Mellin integral transforms and gives
rise to a Mellin-Barnes integral. The result is that the S\'ersic
luminosity density can be written compactly in terms of a Fox $H$
function, which reduces to a Meijer $G$ function for all rational
values of $m$.

The obvious similarity between these two cases invites us to apply
the same Mellin transform technique \citep{0853125287,1996MER.16..05,159829184X}
to the integral (\ref{eq:Sigma_integral}). The basic idea behind
this technique is that any definite integral 
\begin{equation}
  f(z) = \int_{0}^{\infty}g(t,z)\,{\text{d}}t,\label{Marichev}
\end{equation}
can be written as 
\begin{equation}
  f(z) = \int_{0}^{\infty}f_{1}(t)\, f_{2}\left(\frac{z}{t}\right)\,\frac{{\text{d}}t}{t}.
\end{equation}
This expression is exactly a Mellin convolution of two functions
$f_{1}$ and $f_{2}$. Now we can apply the Mellin convolution theorem,
which states that the Mellin transform of a Mellin convolution is
equal to the products of the Mellin transforms of the original
functions.  As a result, the definite integral (\ref{Marichev}) can be
written as the inverse Mellin transform of the product of the
Mellin transforms of $f_{1}$ and $f_{2}$. The Mellin transform
and its inverse are defined as
\begin{gather}
  {\mathfrak{M}}_{f}(u)=\phi(u)=\int_{0}^{\infty}f(z)\, z^{u-1}\,{\text{d}}z,\\
  {\mathfrak{M}}_{\phi}^{-1}(z)=f(z)=\frac{1}{2\pi i}\int_{\mathcal{L}}\phi(u)\, z^{-u}\,{\text{d}}u,
\end{gather}
with ${\mathcal{L}}$ a line integral over a vertical line in the
complex plane. This implies that 
\begin{equation}
  f(z) = \frac{1}{2\pi i}\int_{\mathcal{L}}{\mathfrak{M}}_{f_{1}}(u)\,{\mathfrak{M}}_{f_{2}}(u)\, z^{-u}\,{\text{d}}u.\label{MellinBarnes}
\end{equation}
It now turns out that the integral (\ref{MellinBarnes}) is a
Mellin-Barnes integral for large classes of the functions $f_{1}$ and
$f_{2}$, and that this integral can be evaluated as a Fox $H$ function
or a Meijer $G$ function in many cases.

We can immediately apply this formalism to the integral (\ref{eq:Sigma_integral}),
with $z=1$ and 
\begin{equation}
  f_{1}(t) = 2\,\rho_{0}\, h\,{\text{e}}^{-t^{1/n}}t^{2},
\end{equation}
and 
\begin{equation}
  f_{2}(t) = \begin{cases}
\;\dfrac{t}{\sqrt{1-x^{2}t^{2}}} & \qquad\text{if }0\leq t\leq x^{-1},\\
\;0 & \qquad\text{else}.
\end{cases}
\end{equation}
The Mellin transforms of these functions are readily calculated 
\begin{gather}
  {\mathfrak{M}}_{f_{1}}(u)=2\,\rho_{0}\, h\, n\,\Gamma\,[n(2+u)],\\
  {\mathfrak{M}}_{f_{2}}(u)=\frac{\sqrt{\pi}\,\Gamma\left(\frac{1+u}{2}\right)}{\Gamma\left(\frac{u}{2}\right)}\,\frac{1}{u\, x^{1+u}}.
\end{gather}
Combining these results, we obtain 
\begin{multline}
  \Sigma(x) = 2n\,\sqrt{\pi}\,\rho_{0}\, h\, x\,\\
  \times\,\frac{1}{2\pi i}\int_{\mathcal{L}}\frac{\Gamma\left(2ny\right)\Gamma\left(-\frac{1}{2}+y\right)}{\Gamma\left(y\right)}\left[\, x^{2}\right]^{-y}{\text{d}}y.\label{eq:Sigma-contour}
\end{multline}
This Mellin-Barnes integral may be recognized as a Fox $H$ function,
which is generally defined as the inverse Mellin transform of a product
of gamma-functions, 
\begin{multline}
  H_{p,q}^{m,n}\left[\left.\begin{matrix}({\boldsymbol{a}},{\boldsymbol{A}})\\
        ({\boldsymbol{b}},{\boldsymbol{B}})
      \end{matrix}\,\right|\, z\right]=\\
  \frac{1}{2\pi i}\int_{{\cal L}}\frac{\prod_{j=1}^{m}\Gamma(b_{j}+B_{j}s)\prod_{j=1}^{n}\Gamma(1-a_{j}-A_{j}s)}{\prod_{j=m+1}^{q}\Gamma(1-b_{j}-B_{j}s)\prod_{j=n+1}^{p}\Gamma(a_{j}+A_{j}s)}\, z^{-s}\,{\text{d}}s.\label{eq:defH}
\end{multline}
Using this definition, we can write the surface density of the Einasto
model in the following compact form,
\begin{equation}
  \Sigma\left(x\right)=2n\,\sqrt{\pi}\,\rho_{0}\, h\, x\, H_{1,2}^{2,0}\left[\begin{array}{c}
      (0,\,1)\\
      (0,\,2n),(-\frac{1}{2},\,1)
\end{array}\biggr|\, x^{2}\right].\label{eq:Sigma-Fox}
\end{equation}
The Fox $H$ function is a general analytical function that is becoming
more and more used both in mathematics and applied sciences; in fact,
it is scheduled for inclusion in the \texttt{{Mathematica}} numerical
library. While not the most mainstream special function, the Fox $H$
is an extremely flexible function that contains very broad classes of
elementary and special functions as particular cases.  It is indeed a
very powerful tool for analytical work. It has many general properties
that allow one to manipulate expressions to equivalent forms, reduce the
order for certain values of the parameters, etc.  For details on its
many useful properties, we refer the interested reader to
\citet*{0470263806}, \citet{0415299160}, or \citet{mathai2009h} and the
references therein.

As an illustration of the power of the Fox $H$ function as an analytical
tool, and as a sanity check on the formula, we can calculate the total
mass of the Einasto model by integrating the surface mass density
over the plane of the sky,
\begin{multline}
  M=2\pi\int_{0}^{\infty}\Sigma(\xi)\,\xi\,{\text{d}}\xi=2n\,\pi^{3/2}\,\rho_{0}\, h^{3}\,\\
  \times\ \int_{0}^{\infty}H_{1,2}^{2,0}\left[\begin{array}{c}
      (0,\,1)\\
      (0,\,2n),(-\frac{1}{2},\,1)
    \end{array}\biggr|\, t\right]\, t^{1/2}{\text{d}}t.
\end{multline}
This integral can be calculated by setting $s=\tfrac{3}{2}$
and $a=1$ in equation (2.8) in \citet{mathai2009h}. We immediately
find 
\begin{equation}
  M=2\,\pi\,\rho_{0}\, h^{3}\, n\,\frac{\Gamma\left(3n\right)\,\Gamma(1)}{\Gamma(\frac{3}{2})}=4\pi\,\rho_{0}\, h^{3}\, n\,\Gamma\left(3n\right),
\end{equation}
in agreement with equation (\ref{Mtot}).

An important quantity for gravitational lensing studies is the cumulative
surface mass density, i.e. the total mass contained in a infinite
cylinder with radius $\xi$,
\begin{equation}
  M\left(\xi\right) = 2\pi\int_{0}^{\xi}\Sigma(\xi')\,\xi'\,{\text{d}}\xi'.\label{eq:M-def}
\end{equation}
We find
\begin{multline}
  M(x)=2n\,\pi^{3/2}\,\rho_{0}\, h^{3}\, x^{3}\,\\
  \times\, H_{2,3}^{2,1}\left[\begin{array}{c}
      (-\frac{1}{2},\,1),(0,\,1)\\
      (0,\,2n),(-\frac{1}{2},\,1),(-\frac{3}{2},\,1)
    \end{array}\biggr|\, x^{2}\right].\label{eq:M-Fox}
\end{multline}
As another demonstration of the usefulness of the Fox $H$ function, we
can apply the residue theorem to the contour integral (\ref{eq:defH}),
and obtain explicit series expansions. The general procedure can be
found in \citet{KilbasSaigo99}, and a specific application to the
deprojected S\'ersic model is given in \citet{2011A&A...534A..69B}.
The analysis for the projected Einasto profile is completely
analogous: again, the form of the series expansion depends on the
multiplicity of the poles of the gamma functions
$\Gamma(b_{j}+B_{j}s)$. For both $\Sigma(x)$ and $M(x)$, the poles of
these gamma functions are $-k_{1}/2n$ and $1/2-k_{2}$, with $k_{1}$
and $k_{2}$ any natural number. We encounter two cases:

Case $1$: if $n$ is either non-rational or a rational number $p/q$
with an even denominator (and $p,q$ coprime), all poles are simple and
the expansion is a power series, 
\begin{multline}
  \Sigma(x)=2n\sqrt{\pi}\,\rho_{0}\, h\,\left[\sum_{k=1}^{\infty}\frac{\Gamma\left(-\tfrac{1}{2}-\tfrac{k}{2n}\right)}{\Gamma\left(-\tfrac{k}{2n}\right)}\,\frac{(-1)^{k}}{k!}\,\frac{x^{k/n+1}}{2n}\right.\\
  \left.+\sum_{k=0}^{\infty}\frac{\Gamma(n-2nk)}{\Gamma\left(\tfrac{1}{2}-k\right)}\,\frac{(-1)^{k}}{k!}\, x^{2k}\right],\label{sigma-gen-series}
\end{multline}
and 
\begin{multline}
  M(x)=2n\,\pi^{3/2}\,\rho_{0}\, h^{3}\,\left[-\sum_{k=1}^{\infty}\frac{\Gamma\left(-\tfrac{3}{2}-\tfrac{k}{2n}\right)}{\Gamma\left(-\tfrac{k}{2n}\right)}\,\frac{(-1)^{k}}{k!}\,\frac{x^{k/n+3}}{2n}\right.\\
  \left.+\sum_{k=0}^{\infty}\frac{\Gamma(n-2nk)}{\Gamma\left(\tfrac{1}{2}-k\right)}\,\frac{(-1)^{k}}{(k+1)!}\, x^{2k+2}\right].\label{mass-gen-series}
\end{multline}
Case $2$: if $n$ is integer or a rational number $p/q$
with an odd denominator, some poles are of second order, and the expansion
is a logarithmic-power series. If we define $k_{0}=\tfrac{q-1}{2}$,
then we obtain after some algebra
\begin{multline}
  \Sigma(x)=2n\sqrt{\pi}\,\rho_{0}\, h\,\left[\;\sum_{\substack{k=1\\
k\,{\text{mod}}\, p\ne0
}
}^{\infty}\hspace{-1.5ex}\frac{\Gamma\left(-\tfrac{1}{2}-\tfrac{k}{2n}\right)}{\Gamma\left(-\tfrac{k}{2n}\right)}\,\frac{(-1)^{k}}{k!}\,\frac{x^{k/n+1}}{2n}\right.\\
\left.\ +\ \frac{\Gamma(n)}{\sqrt{\pi}}\ +\ \hspace{-3ex}\sum_{\substack{k=1\\
(k+k_{0})\,{\text{mod}}\, q\ne0
}
}^{\infty}\hspace{-3ex}\frac{\Gamma(n-2nk)}{\Gamma\left(\tfrac{1}{2}-k\right)}\,\frac{(-1)^{k}}{k!}\, x^{2k}\right]\\
\ +\ 2\,\rho_{0}\, h\,\hspace{-3ex}\sum_{\substack{k=0\\
(k+k_{0})\,{\text{mod}}\, q=0
}
}^{\infty}\hspace{-3ex}\frac{(-1)^{p}\,(2k)!}{(2nk-n)!\, k!\, k!}\,\left(\frac{x}{2}\right)^{2k}\ \times\left[-\ln\left(\frac{x}{2}\right)\right.\\
\left.-\frac{1}{2k}+\psi(k+1)+n\,\psi(2nk-n)-\psi(2k-1)\right],\label{sigma-rat-series}
\end{multline}
and 
\begin{multline}
  M(x)=2n\,\pi^{3/2}\,\rho_{0}\, h^{3}\, x^{2}\,\left[\ \ -\hspace{-1.5ex}\sum_{\substack{k=1\\
k\,{\text{mod}}\, p\ne0
}
}^{\infty}\hspace{-1ex}\frac{\Gamma\left(-\tfrac{3}{2}-\tfrac{k}{2n}\right)}{\Gamma\left(-\tfrac{k}{2n}\right)}\,\frac{(-1)^{k}}{k!}\,\frac{x^{k/n+1}}{2n}\right.\\
\left.+\ \frac{\Gamma(n)}{\sqrt{\pi}}\ +\ \hspace{-3ex}\sum_{\substack{k=1\\
(k+k_{0})\,{\text{mod}}\, q\ne0
}
}^{\infty}\hspace{-3ex}\frac{\Gamma(n-2nk)}{\Gamma\left(\tfrac{1}{2}-k\right)}\,\frac{(-1)^{k}}{(k+1)!}\, x^{2k}\right]\\
\ +\ 2\pi\,\rho_{0}\, h^{3}\, x^{2}\,\hspace{-3ex}\sum_{\substack{k=0\\
(k+k_{0})\,{\text{mod}}\, q=0
}
}^{\infty}\hspace{-3ex}\frac{(-1)^{p}\,(2k)!}{(2nk-n)!\, k!\,(k+1)!}\left(\frac{x}{2}\right)^{2k}\times\ \left[-\ln\left(\frac{x}{2}\right)\right.\\
\left.-\frac{1}{2k}-\frac{1}{2k+2}+\psi(k+2)+n\,\psi(2nk-n)-\psi(2k-1)\right],\label{mass-rat-series}
\end{multline}
with $\psi(k)$ the digamma function.

\subsection{Lens equation and deflection angle }

In the thin lens approximation, the lens equation for axially symmetric
lens is 
\begin{equation}
  \eta = \frac{D_{S}}{D_{L}}\,\xi-D_{LS}\,\hat{\alpha},\label{eq:lens-equation-dimensional}
\end{equation}
where the quantities $\eta$ and $\xi$ are the physical positions
of the source in the source plane and an image in the image plane,
respectively, $\hat{\alpha}$ is the deflection angle, and $D_{L}$,
$D_{S}$ and $D_{LS}$ are the angular distances from observer to
lens, from observer to source, and from lens to source, respectively.

With the dimensionless positions $y=D_{L}\,\eta/D_{S}\, h$ and $x=\xi/h$,
and dimensionless $\alpha=D_{L}\, D_{LS}\hat{\alpha}/D_{S}\,\xi$,
the lens equation reduces to
\begin{equation}
  y = x-\alpha(x).\label{eq:lens-equation-adimensional}
\end{equation}
The deflection angle for a spherical symmetric lens is \citep{1992grle.book.....S}
\begin{equation}
  \alpha(x)=\frac{2}{x}\int_{0}^{x}x'\,\frac{\Sigma(x')}{\Sigma_{{\text{crit}}}}\,{\text{d}}x'=\frac{2}{x}\int_{0}^{x}x'\,\kappa(x')\,{\text{d}}x',\label{eq:alpha}
\end{equation}
where 
\begin{equation}
  \kappa = \frac{\Sigma(x)}{\Sigma_{{\text{crit}}}},\label{eq:convergence}
\end{equation}
is the convergence and $\Sigma_{{\text{crit}}}$ is the critical
surface mass density defined by 
\begin{equation}
  \Sigma_{{\text{crit}}} \equiv \frac{c^{2}\, D_{\text{S}}}{4\pi\, G\, D_{\text{L}}\, D_{\text{LS}}},
\end{equation}
where $c$ is the speed of light, and $G$ is the gravitational constant.
Evidently, the deflection angle is related to the integrated mass
as
\begin{equation}
  \alpha(x) = \frac{M(x)}{\pi\, h^{2}\,\Sigma_{{\text{crit}}}x}.
\end{equation}
Introducing the central convergence, $\kappa_{\text{c}}$, a parameter
that determines the lensing properties of the Einasto profile, 
\begin{equation}
  \kappa_{\text{c}}\equiv\frac{\Sigma\left(0\right)}{\Sigma_{{\text{crit}}}}=\frac{2\,\rho_{0}\, h\, n\,\Gamma\left(n\right)}{\Sigma_{{\text{crit}}}},
\end{equation}
we can write $\alpha\left(x\right)$ in the form
\begin{equation}
  \alpha\left(x\right)=\frac{\kappa_{\text{c}}\,\sqrt{\pi}}{\Gamma\left(n\right)}\, x^{2}\, H_{2,3}^{2,1}\left[\begin{array}{c}
(-\frac{1}{2},\,1),(0,\,1)\\
(0,\,2n),(-\frac{1}{2},\,1),(-\frac{3}{2},\,1)
\end{array}\biggr|\, x^{2}\right],\label{eq:alpha-Fox}
\end{equation}
with a completely analogous series expansion as for $M(x)$. For a
spherically symmetric lens that is capable of forming multiple images of
the source, one sufficient condition is $\kappa_{\text{c}}>1$
\citep{1992grle.book.....S}. In the case $\kappa_{\text{c}}\leq1$,
only one image of the source is formed. In addition to the condition
$\kappa_{\text{c}}>1$, multiple images are produced only if $\mid
y\mid\leq y_{{\text{crit}}}$ \citep{2002ApJ...566..652L}, where
$y_{{\text{crit}}}$ is the maximum value of $y$ when $x<0$ or its
minimum when $x>0$. For singular profiles such as an NFW profile, the
central convergence always is divergent, hence the condition
$\kappa_{\text{c}}>1$ is always met, this implies that an NFW profile is
capable of forming multiple images for any mass.  Nonsingular
profiles, such as the Einasto profile, are not capable of forming
multiple images for any mass. Instead, the condition
$\kappa_{\text{c}}>1$ sets a minimum value for lens mass required to
form multiple images.

\subsection{Deflection potential}

The deflection potential $\psi\left(x\right)$ for a spherically symmetric
lens is given by \citep{1992grle.book.....S} 
\begin{equation}
\psi(x)=2\int_{0}^{x}x'\,\kappa(x')\,\ln\left(\dfrac{x}{x'}\right)\,{\text{d}}x'.\label{eq:phi}
\end{equation}
Inserting eq.~(\ref{eq:Sigma-Fox}) into (\ref{eq:phi}),
we obtain again a result that can be re-expressed as a Fox $H$ function
\begin{multline}
\psi(x)=\frac{\kappa_{\text{c}}\,\sqrt{\pi}}{2\,\Gamma\left(n\right)}\, x^{3}\,\\
\times\, H_{3,4}^{2,2}\left[\begin{array}{c}
(-\frac{1}{2},\,1),(-\frac{1}{2},\,1),(0,\,1)\\
(0,\,2n),(-\frac{1}{2},\,1),(-\frac{3}{2},\,1),(-\frac{3}{2},\,1)
\end{array}\biggr|\, x^{2}\right].\label{eq:phi-Fox}
\end{multline}
For $\psi(x)$ we have the same cases of expansions, a power series
for simple poles in case $1$, 
\begin{multline}
\psi(x)=\frac{\kappa_{\text{c}}\,\sqrt{\pi}}{2\,\Gamma\left(n\right)}\,\left[-\sum_{k=1}^{\infty}\frac{\Gamma\left(-\tfrac{3}{2}-\tfrac{k}{2n}\right)}{\Gamma\left(-\tfrac{k}{2n}\right)}\,\frac{(-1)^{k}}{k!}\,\frac{x^{k/n+3}}{3n+k}\right.\\
\left.+\sum_{k=0}^{\infty}\frac{\Gamma(n-2nk)}{\Gamma\left(\tfrac{1}{2}-k\right)}\,\frac{(-1)^{k}}{(k+1)!}\,\frac{x^{2k+2}}{k+1}\right],\label{eq:phi-gen-series}
\end{multline}
and a logarithmic-power series for multiple poles in case $2$, 
\begin{multline}
\psi(x)=\frac{\kappa_{\text{c}}\,\sqrt{\pi}}{2\,\Gamma\left(n\right)}\, x^{2}\,\left[\ \ -\hspace{-1.5ex}\sum_{\substack{k=1\\
k\,{\text{mod}}\, p\ne0
}
}^{\infty}\hspace{-1ex}\frac{\Gamma\left(-\tfrac{3}{2}-\tfrac{k}{2n}\right)}{\Gamma\left(-\tfrac{k}{2n}\right)}\,\frac{(-1)^{k}}{k!}\,\frac{x^{k/n+1}}{3n+k}\right.\\
\left.+\ \frac{\Gamma(n)}{\sqrt{\pi}}\ +\ \hspace{-3ex}\sum_{\substack{k=1\\
(k+k_{0})\,{\text{mod}}\, q\ne0
}
}^{\infty}\hspace{-3ex}\frac{\Gamma(n-2nk)}{\Gamma\left(\tfrac{1}{2}-k\right)}\,\frac{(-1)^{k}}{(k+1)!}\,\frac{x^{2k}}{k+1}\right]\\
\ +\ \frac{\kappa_{\text{c}}}{2\,\Gamma\left(n\right)}\, x^{2}\,\hspace{-3ex}\sum_{\substack{k=0\\
(k+k_{0})\,{\text{mod}}\, q=0
}
}^{\infty}\hspace{-3ex}\frac{(-1)^{p}\,(2k)!}{(2nk-n)!\,(k+1)!\,(k+1)!}\left(\frac{x}{2}\right)^{2k}\times\ \\
\left.\left[-\ln\left(\frac{x}{2}\right)\right.-\frac{1}{2k}+\psi(k+2)+n\,\psi(2nk-n)-\psi(2k-1)\right].\label{eq:phi-rat-series}
\end{multline}

\subsection{Magnification, shear and the critical curves}

Gravitational lensing effect preserves the surface brightness but
it causes variations in shape and solid angle of the source. Therefore,
the source luminosity is amplified by \citep{1992grle.book.....S}
\begin{equation}
  \mu=\frac{1}{\left(1-\kappa\right)^{2}-\gamma^{2}},\label{eq:magnification}
\end{equation}
where $\kappa$ is the convergence and $\gamma=\gamma\left(x\right)$
is the shear. The amplification $\mu$ has two contributions: one
from the convergence, which describes an isotropic focusing of light
rays in the lens plane, and the other which describes an anisotropic
focusing caused by the tidal gravitational forces acting on the light
rays, described by the shear. For a spherical symmetric lens, the
shear is given by \citep{1991ApJ...370....1M}
\begin{equation}
  \gamma\left(x\right)=\frac{\bar{\Sigma}\left(x\right)-\Sigma\left(x\right)}{\Sigma_{crit}}=\bar{\kappa}-\kappa,\label{eq:shear}
\end{equation}
where the average surface mass density within $x$ is 
\begin{equation}
  \bar{\Sigma}(x)=\frac{2}{x^{2}}\int_{0}^{x}x'\,\Sigma(x')\,{\text{d}}x'.\label{eq:mean-sigma}
\end{equation}
The magnification of the Einasto profile can be found combining
eqs.~(\ref{eq:Sigma-Fox}), (\ref{eq:magnification}), (\ref{eq:shear})
and (\ref{eq:mean-sigma}). In the calculation of
$\bar{\Sigma}\left(x\right)$, we use again the Mellin-Barnes integral
representation (\ref{eq:Sigma-contour}), and integrate it to obtain an
expression in terms of Fox $H$ function.  Thus we get
\begin{equation}
\mu=\left[\left(1-\bar{\kappa}\right)\left(1+\bar{\kappa}-2\kappa\right)\right]^{-1},\label{eq:magnification-einasto}
\end{equation}
where
\begin{equation}
  \kappa\left(x\right)=\frac{\kappa_{\text{c}}\,\sqrt{\pi}}{\Gamma\left(n\right)}\, x\, H_{1,2}^{2,0}\left[\begin{array}{c}
(0,\,1)\\
(0,\,2n),(-\frac{1}{2},\,1)
\end{array}\biggr|\, x^{2}\right],\label{eq:kappa-Fox}
\end{equation}
and
\begin{equation}
  \bar{\kappa}\left(x\right)=\frac{\kappa_{\text{c}}\,\sqrt{\pi}}{\Gamma\left(n\right)}\, x\, H_{2,3}^{2,1}\left[\begin{array}{c}
(-\frac{1}{2},\,1),(0,\,1)\\
(0,\,2n),(-\frac{1}{2},\,1),(-\frac{3}{2},\,1)
\end{array}\biggr|\, x^{2}\right].\label{eq:average-kappa-Fox}
\end{equation}
The magnification may be divergent for some image positions.  The loci
of the diverging magnification in the image plane are called the
critical curves. We see from eq.~(\ref{eq:magnification-einasto}) that the Einasto
profile has one pair of critical
curves. The first curve, $1-\bar{\kappa}=0$, is the tangential
critical curve, which corresponds to an Einstein Ring with a radius,
called the Einstein radius. The second curve,
$1+\bar{\kappa}-2\kappa=0$, is the radial critical curve, which also
defines a ring and its corresponding radius. In both cases the
equations must be solved numerically.

\section{Integer and half-integer values of $n$ \label{sec:04}}

The expressions for the Einasto surface mass density (\ref{eq:Sigma-Fox})
and its lensing properties (\ref{eq:M-Fox}), (\ref{eq:alpha-Fox}), (\ref{eq:phi-Fox}),
(\ref{eq:magnification-einasto}), (\ref{eq:kappa-Fox}) and (\ref{eq:average-kappa-Fox})
in terms of the Fox $H$ function for case $n$ is a rational number,
can be reduced to a Meijer $G$ function,
\begin{multline}
G_{p,q}^{m,n}\left[\left.\begin{matrix}{\boldsymbol{a}}\\
{\boldsymbol{b}}
\end{matrix}\,\right|\, z\right]=\\
\frac{1}{2\pi i}\int_{{\cal \mathcal{L}}}\frac{\prod_{j=1}^{m}\Gamma(b_{j}+s)\prod_{j=1}^{n}\Gamma(1-a_{j}-s)}{\prod_{j=m+1}^{q}\Gamma(1-b_{j}-s)\prod_{j=n+1}^{p}\Gamma(a_{j}+s)}\, z^{-s}\,{\text{d}}s.\label{eq:defG}
\end{multline}
Meijer $G$ function numerical routines had been implemented
in computer algebraic systems, such as the commercial \texttt{{Maple}},
\texttt{{Mathematica}} and the free open-source \texttt{{Sage}}
and \texttt{{mpmath}} library; in contrast, there is no Fox $H$
function implementation available. 

For case $n$ integer or half-integer, we can write $\Sigma\left(x\right)$
as a Meijer $G$ function. Inserting the Gauss multiplication formula
\citep{1970hmfw.book.....A}
\begin{equation}
  \Gamma(2ny)=(2n)^{-\frac{1}{2}+2ny}\,(2\pi)^{\frac{1}{2}-n}\,\Gamma(y)\,\prod_{j=1}^{2n-1}\Gamma\left(\frac{j}{2n}+y\right),\label{eq:gauss-mult-form}
\end{equation}
into eq.~(\ref{eq:Sigma-contour}) and comparing with the
definition (\ref{eq:defG}), we obtain an expression for the surface
mass density of the Einasto profile in terms of the Meijer $G$ function
\begin{subequations}
\label{Sigma-Meijer} 
\begin{equation}
  \Sigma(x)=\frac{\sqrt{n}\,\rho_{0}\, h}{(2\pi)^{n-1}}\, x\, G_{0,2n}^{2n,0}\left[\begin{array}{c}
-\\
{\boldsymbol{b}}
\end{array}\biggr|\,\frac{x^{2}}{\left(2n\right)^{2n}}\right],
\end{equation}
where ${\boldsymbol{b}}$ is a vector of size $2n$ given by
\begin{equation}
  {\boldsymbol{b}}=\biggl\{\frac{1}{2n},\frac{2}{2n},\ldots,\frac{2n-1}{2n},-\frac{1}{2}\biggr\}.
\end{equation}
\end{subequations} 
Integrating the above expression for surface mass density according to
(\ref{eq:M-def}) and using the integral properties of the Meijer $G$
function (eq.~07.34.21.0003.01 on the Wolfram Functions Site%
\footnote{http://functions.wolfram.com/HypergeometricFunctions/MeijerG/%
}), we obtain
\begin{equation}
  M\left(x\right)=\frac{\sqrt{n}\,\rho_{0}\, h^{3}}{2(2\pi)^{n-2}}\, x^{3}\, G_{1,2n+1}^{2n,1}\left[\begin{array}{c}
-\frac{1}{2}\\
{\boldsymbol{b}},-\frac{3}{2}
\end{array}\biggr|\,\frac{x^{2}}{\left(2n\right)^{2n}}\right].\label{eq:M-Meijer}
\end{equation}
Inserting eq.~(\ref{Sigma-Meijer}) into (\ref{eq:alpha})
and performing the integral of the Meijer $G$ function, we find
\begin{equation}
  \alpha\left(x\right)=\frac{\kappa_{\text{c}}}{2\,(2\pi)^{n-1}\,\sqrt{n}\,\Gamma\left(n\right)}\, x^{2}\, G_{1,2n+1}^{2n,1}\left[\begin{array}{c}
-\frac{1}{2}\\
{\boldsymbol{b}},-\frac{3}{2}
\end{array}\biggr|\,\frac{x^{2}}{\left(2n\right)^{2n}}\right].\label{eq:alpha-Meijer}
\end{equation}
We obtain the deflection potential in terms of the Meijer $G$ function
combining eqs.~(\ref{eq:phi}),~(\ref{Sigma-Meijer}) and again we
integrate, applying the Meijer $G$ function properties.  We obtain
\begin{equation}
  \psi\left(x\right)=\frac{\kappa_{\text{c}}}{4\,(2\pi)^{n-1}\,\sqrt{n}\,\Gamma\left(n\right)}\, x^{3}\, G_{2,2n+2}^{2n,2}\left[\begin{array}{c}
-\frac{1}{2},-\frac{1}{2}\\
{\boldsymbol{b}},-\frac{3}{2},-\frac{3}{2}
\end{array}\biggr|\,\frac{x^{2}}{\left(2n\right)^{2n}}\right].\label{eq:phi-Meijer}
\end{equation}
The convergence can be simply found dividing eq.~(\ref{Sigma-Meijer})
by $\Sigma_{\text{crit}}$
\begin{equation}
  \kappa\left(x\right)=\frac{\kappa_{\text{c}}}{2\,(2\pi)^{n-1}\,\sqrt{n}\,\Gamma\left(n\right)}\, x\, G_{0,2n}^{2n,0}\left[\begin{array}{c}
-\\
{\boldsymbol{b}}
\end{array}\biggr|\,\frac{x^{2}}{\left(2n\right)^{2n}}\right],\label{eq:kappa-Meijer}
\end{equation}
the average convergence can be calculated by inserting
eq.~(\ref{Sigma-Meijer}) into eq.~(\ref{eq:mean-sigma}), and then
integrating, we find
\begin{equation}
  \bar{\kappa}\left(x\right)=\frac{\kappa_{\text{c}}}{2\,(2\pi)^{n-1}\,\sqrt{n}\,\Gamma\left(n\right)}\, x\, G_{1,2n+1}^{2n,1}\left[\begin{array}{c}
-\frac{1}{2}\\
{\boldsymbol{b}},-\frac{3}{2}
\end{array}\biggr|\,\frac{x^{2}}{\left(2n\right)^{2n}}\right].\label{eq:mean-kappa-Meijer}
\end{equation}

\subsection{Simple cases: $n=1$ and $n=\frac{1}{2}$ }

The Einasto profile corresponding to $n=1$ is a simple exponential
model, characterised by the density profile 
\begin{equation}
\rho\left(r\right)=\rho_{0}\,\exp\left\{ -\left(\frac{r}{h}\right)\,\right\} ,\label{rho-n1}
\end{equation}
the surface density profile and lensing properties are readily found
by setting $n=1$ in eqs.~(\ref{Sigma-Meijer}), (\ref{eq:M-Meijer}),
(\ref{eq:alpha-Meijer}), (\ref{eq:phi-Meijer}), (\ref{eq:kappa-Meijer})
and (\ref{eq:mean-kappa-Meijer}) 
\begin{gather}
  \Sigma(x)=\rho_{0}\, h\, x\, G_{0,2}^{2,0}\left[\begin{array}{c}
    -\\
    \frac{1}{2},-\frac{1}{2}
  \end{array}\biggr|\,\frac{x^{2}}{4}\right],\\
M\left(x\right)=\pi\,\rho_{0}\, h^{3}\, x^{3}\, G_{1,3}^{2,1}\left[\begin{array}{c}
    -\frac{1}{2}\\
    \frac{1}{2},-\frac{1}{2},-\frac{3}{2}
  \end{array}\biggr|\,\frac{x^{2}}{4}\right],\\
\alpha\left(x\right)=\frac{\kappa_{\text{c}}}{2}\, x^{2}\, G_{1,3}^{2,1}\left[\begin{array}{c}
    -\frac{1}{2}\\
    \frac{1}{2},-\frac{1}{2},-\frac{3}{2}
  \end{array}\biggr|\,\frac{x^{2}}{4}\right],\\
\psi\left(x\right)=\frac{\kappa_{\text{c}}}{4}\, x^{3}\, G_{2,4}^{2,2}\left[\begin{array}{c}
    -\frac{1}{2},-\frac{1}{2}\\
    \frac{1}{2},-\frac{1}{2},-\frac{3}{2},-\frac{3}{2}
  \end{array}\biggr|\,\frac{x^{2}}{4}\right],\\
\kappa\left(x\right)=\frac{\kappa_{\text{c}}}{2}\, x\, G_{0,2}^{2,0}\left[\begin{array}{c}
    -\\
    \frac{1}{2},-\frac{1}{2}
  \end{array}\biggr|\,\frac{x^{2}}{4}\right],\\
\bar{\kappa}\left(x\right)=\frac{\kappa_{\text{c}}}{2}\, x\, G_{1,3}^{2,1}\left[\begin{array}{c}
    -\frac{1}{2}\\
    \frac{1}{2},-\frac{1}{2},-\frac{3}{2}
  \end{array}\biggr|\,\frac{x^{2}}{4}\right].
\end{gather}
The specific Meijer $G$ functions in these expressions can be reduced
to more standard special functions. We obtain the equivalent expressions
\begin{gather}
  \Sigma\left(x\right)=2\rho_{0}\, h\, x\, K_{1}\left(x\right),\label{Sigma-n1}\\
  M\left(x\right)=8\pi\,\rho_{0}\, h^{3}\,\left[1-\frac{x^{2}}{2}\, K_{2}\left(x\right)\right],\label{eq:M-n1}\\
  \alpha\left(x\right)=\frac{4\kappa_{\text{c}}}{x}\,\left[1-\frac{x^{2}}{2}\, K_{2}\left(x\right)\right],\label{alpha-n1}\\
  \psi\left(x\right)=4\kappa_{\text{c}}\left[\ln\left(\frac{x}{2}\right)+\frac{x}{2}\, K_{1}\left(x\right)+K_{0}\left(x\right)+\gamma-\frac{1}{2}\right],\label{eq:phi-n1}\\
  \kappa\left(x\right)=\kappa_{\text{c}}\, x\, K_{1}\left(x\right),\label{eq:kappa-n1}\\
  \bar{\kappa}\left(x\right)=\frac{4\kappa_{\text{c}}}{x^{2}}\,\left[1-\frac{x^{2}}{2}\,
    K_{2}\left(x\right)\right],\label{eq:mean-kappa-n1}
\end{gather}
where $K_{\nu}(x)$ is the modified Bessel function of the second kind
of order $\nu$, and $\gamma\approx0.57721566$ the Euler-Mascheroni
constant. The expressions (\ref{Sigma-n1}), (\ref{eq:M-n1}), (\ref{alpha-n1}),
(\ref{eq:phi-n1}), (\ref{eq:kappa-n1}) and (\ref{eq:mean-kappa-n1})
can also be readily calculated by inserting the density profile (\ref{rho-n1})
into the formula (\ref{eq:04}) and later carrying out all corresponding
calculations for the lensing properties.

For $n=\tfrac{1}{2}$, the density profile (\ref{eq:einasto_halo_vers})
becomes a Gaussian, 
\begin{equation}
  \rho\left(r\right)=\rho_{0}\,\exp\left\{ -\left(\frac{r}{h}\right)^{2}\,\right\} ,\label{eq:pho-n-1/2}
\end{equation}
the surface density and lensing properties can be found through substitution
of $n=\tfrac{1}{2}$ in eqs.~(\ref{Sigma-Meijer}), (\ref{eq:alpha-Meijer}),
(\ref{eq:M-Meijer}), (\ref{eq:phi-Meijer}), (\ref{eq:kappa-Meijer})
and (\ref{eq:mean-kappa-Meijer}). 
\begin{gather}
  \Sigma(x)=\sqrt{\pi}\,\rho_{0}\, h\, x\, G_{0,1}^{1,0}\left[\begin{array}{c}
-\\
-\frac{1}{2}
\end{array}\biggr|\, x^{2}\right],\\
M\left(x\right)=\pi^{3/2}\,\rho_{0}\, h^{3}\, x^{3}\, G_{1,2}^{1,1}\left[\begin{array}{c}
-\frac{1}{2}\\
-\frac{1}{2},-\frac{3}{2}
\end{array}\biggr|\, x^{2}\right],\\
\alpha\left(x\right)=\kappa_{\text{c}}\, x^{2}\, G_{1,2}^{1,1}\left[\begin{array}{c}
-\frac{1}{2}\\
-\frac{1}{2},-\frac{3}{2}
\end{array}\biggr|\, x^{2}\right],\\
\psi\left(x\right)=\frac{\kappa_{\text{c}}}{2}\, x^{3}\, G_{2,3}^{1,2}\left[\begin{array}{c}
-\frac{1}{2},-\frac{1}{2}\\
-\frac{1}{2},-\frac{3}{2},-\frac{3}{2}
\end{array}\biggr|\, x^{2}\right],\\
\kappa\left(x\right)=\kappa_{\text{c}}\, x\, G_{0,1}^{1,0}\left[\begin{array}{c}
-\\
-\frac{1}{2}
\end{array}\biggr|\, x^{2}\right],\\
\bar{\kappa}\left(x\right)=\kappa_{\text{c}}\, x\, G_{1,2}^{1,1}\left[\begin{array}{c}
-\frac{1}{2}\\
-\frac{1}{2},-\frac{3}{2}
\end{array}\biggr|\, x^{2}\right].
\end{gather}
These Meijer $G$ function can also be written in terms of elementary
and special functions, 
\begin{gather}
\Sigma\left(x\right)=\sqrt{\pi}\rho_{0}\, h\,{\text{e}}^{-x^{2}},\label{Sigma-n-1/2}\\
M\left(x\right)=\pi^{3/2}\,\rho_{0}\, h^{3}\,\left(1-{\text{e}}^{-x^{2}}\right),\label{eq:M-n-1/2}\\
\alpha\left(x\right)=\frac{\kappa_{\text{c}}}{x}\,\left(1-{\text{e}}^{-x^{2}}\right),\label{alpha-n-1/2}\\
\psi\left(x\right)=\frac{\kappa_{\text{c}}}{2}\,\left[\ln\left(x^{2}\right)+\mathrm{E}_{1}(x^{2})+\gamma\right],\label{eq:phi-n-1/2}\\
\kappa\left(x\right)=\kappa_{\text{c}}\,{\text{e}}^{-x^{2}},\label{eq:kappa-n-1/2}\\
\bar{\kappa}\left(x\right)=\frac{\kappa_{\text{c}}}{x^{2}}\,\left(1-{\text{e}}^{-x^{2}}\right),\label{eq:mean-kappa-n-1/2}
\end{gather}
with $\mathrm{E}_{\nu}(x)$ the exponential integral of
order $\nu$. The above results can be easily checked  by
substituting eq.~(\ref{eq:pho-n-1/2}) into recipes (\ref{eq:04}),
(\ref{eq:M-def}), (\ref{eq:alpha}), (\ref{eq:convergence}), (\ref{eq:phi})
and (\ref{eq:mean-sigma}).

\subsection{Profile comparison}

\begin{figure}
\centering{}\includegraphics[scale=0.62]{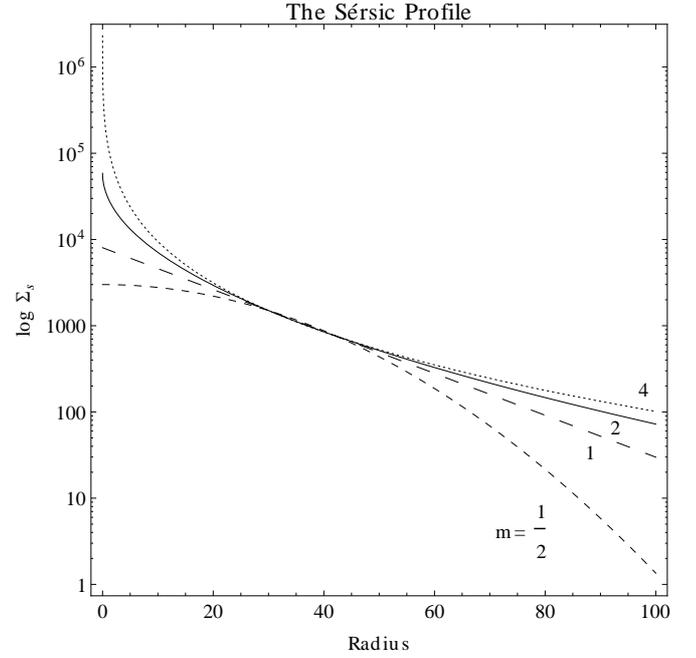}\caption{S\'ersic profile, in dimensionless units, where $\Upsilon I_{{\text{e}}}$
and $R_{{\text{e}}}$ are held fixed for different values of the S\'ersic
index $m$.\label{fig:plot01-Sersic-profile}}
\end{figure}

\begin{figure}
\centering{}\includegraphics[scale=0.62]{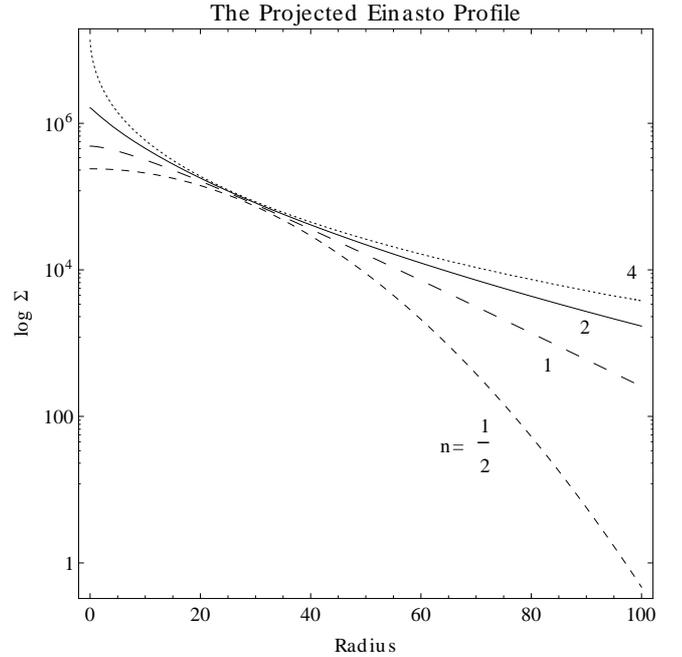}\caption{Projected Einasto profile, where $\rho_{0}{r_{{\text{s}}}}$
and ${r_{{\text{s}}}}$ are held fixed for different values
of the Einasto index $n$.\label{fig:plot02-Einasto-profile}}
\end{figure}

We compared the Einasto and S\'ersic surface mass densities for the
same values of the S\'ersic index $m$ and Einasto index $n$, including
the exponential and Gaussian cases. For this comparison we used the
\texttt{{Mathematica}} implementation of the Meijer $G$ function
and eq.~(\ref{Sigma-Meijer}). Figure \ref{fig:plot01-Sersic-profile}
shows $\Sigma_{{\text{S}}}(R)$ for different values of $m$, while
$\Upsilon I_{{\text{e}}}$ and $R_{{\text{e}}}$ are held fixed; Figure
\ref{fig:plot02-Einasto-profile} displays $\Sigma(x)$ for different
values of $n$, while $\rho_{0}{r_{{\text{s}}}}$ and ${r_{{\text{s}}}}$
are held fixed. In both figures, it can be clearly seen that the respective
index is very important in determining the overall behaviour of the
curves. 

The S\'ersic profile is characterised by a steeper central
core and extended external wing for higher values of the S\'ersic
index $m$. For low values of $m$ the central core is flatter and
the external wing is sharply truncated. The Einasto profile has a
similar behaviour, with the difference that the external wings are
more spread out. Also in the inner region for both profiles with low
values of the respectively index we obtain higher values of $\Sigma_{\text{S}}$
and $\Sigma$. Additionally, the Einasto profile has higher values
of the central surface mass density than the S\'ersic profile, comparing
both profiles for the same index. However, the Einasto profile seems
to be less sensitive to the value of the surface mass density for
a given $n$ and radius in the inner region than the S\'ersic profile.
It is in this region that the lensing effect is more important and
the surface mass density characteristics determine the lensing properties
of the respective profiles. 

Given these differences between the two profiles, we clearly see that
the lensing properties of S\'ersic and Einasto profiles also differ,
which agrees with previous work of \citet{2005MNRAS.358.1325C} and
\citet{2010MNRAS.405..340D}. Studies of the lensing properties of the
S\'ersic profile have been made by \citet{2004A&A...415..839C} and
\citet{2007JCAP...07..006E}.

\section{Asymptotic behaviour \label{sec:05}}

The series expansions allow us to directly investigate the behaviour
of Einasto models at small radii. It follows that the central asymptotic
behaviour of the surface density $\Sigma(x)$ depends on the value
of $n$. If $n<1$, we find the following expansion at small radii
$(x\ll1)$ 
\begin{equation}
  \Sigma(x)\sim\rho_{0}\, h\left[\,2\Gamma(n+1)-\Gamma(1-n)\, x^{2}\,\right].
\end{equation}
If $n=1$, then the expansion has the form 
\begin{equation}
  \Sigma(x)\sim\rho_{0}\, h\left[\,2+\left(2\ln\left(\tfrac{1}{2}\right)-1\right)\, x^{2}\,\right].
\end{equation}
Finally, if $n>1$, the central surface density behaves as 
\begin{equation}
  \Sigma(x)\sim\rho_{0}\, h\left[\,2\Gamma(n+1)-\frac{\sqrt{\pi}}{n+1}\,\frac{\Gamma\left(\tfrac{n-1}{2n}\right)}{\Gamma\left(\tfrac{2n-1}{2n}\right)}\, x^{1+\frac{1}{n}}\,\right].
\end{equation}
The behaviour of the cumulative surface mass $M(x)$ and
deflection potential $\psi\left(x\right)$ are more straightforward.
At small radii, we find the asymptotic expressions 
\begin{equation}
  M(x)\sim2\pi\,\rho_{0}\, h^{3}\,\Gamma(n+1)\, x^{2},\label{M-small}
\end{equation}
and
\begin{equation}
\psi(x)\sim\frac{\kappa_{\text{c}}}{2}\, x^{2}.\label{Phi-small}
\end{equation}
The $x^{2}$ slope for $M(x)$ is not unexpected, given
that the Einasto models have a finite central surface mass density.
The asymptotic behaviour of the Fox $H$ function at large radii is
described in \citet{KilbasSaigo99}. We obtain the following expansions
$(x\gg1)$
\begin{gather}
  \Sigma(x)\sim\sqrt{8n\,\pi}\,\rho_{0}\, h\,{\text{e}}^{-x^{1/n}}\, x^{1-\frac{1}{2n}},\label{Sigma-big}\\
  M(x)\sim4\pi\,\rho_{0}\, h^{3}\, n\,\Gamma(3n)\,-\,2(2\pi n)^{3/2}\,\rho_{0}\, h^{3}\,{\text{e}}^{-x^{1/n}}\, x^{3-\frac{3}{2n}},
\end{gather} 
and
\begin{multline}
  \psi(x)\sim2\,\frac{\Gamma\left(3n\right)}{\Gamma\left(n\right)}\,\kappa_{\text{c}}\,\left[\,\ln\left(\frac{x}{2}\right)-n\,\psi\left(3n\right)+1\,\right]\\
+\frac{\sqrt{\pi}\,\left(2n\right)^{3/2}}{\Gamma\left(n\right)}\,\kappa_{\text{c}}\,{\text{e}}^{-x^{1/n}}\, x^{3-\frac{5}{2n}}.\label{eq:Phi-large}
\end{multline}

\section{Summary and conclusions \label{sec:06}}

We studied the spatial and lensing properties of the Einasto profile
by analytical means. For the spatial properties we applied the method
used by \citet{1999A&A...352..447C} to derive an analytical expansion
for the dimensionless parameter $d_{n}$ of the Einasto model. We
also derived analytical expressions for the cumulative mass profile
$M\left(r\right)$ and the gravitational potential
$\Psi\left(r\right)$.  For the lensing properties, we used the Mellin
integral transform formalism to derive closed, analytical expressions
for the surface mass density $\Sigma\left(x\right)$, cumulative
surface mass $M\left(x\right)$, deflection angle
$\alpha\left(x\right)$, deflection potential $\psi\left(x\right)$,
magnification $\mu\left(x\right)$ and shear $\gamma\left(x\right).$
For general values of the Einasto index $n$, these are expressed in
terms of the Fox $H$ function. Using the properties of the Fox $H$
function, we calculated explicit power and logarithmic-power
expansions for these lensing quantities, and we obtained simplified
expressions for integer and half-integer values of $n$. These series
expansions allow to perform arbitrary-precision calculations of
the surface mass density and lensing properties of Einasto dark matter
haloes. We also studied the asymptotic behaviour of the surface mass
density, cumulative surface mass and deflection potential at small and
large radii using the series expansions.

Furthermore, we compared the S\'ersic and Einasto surface mass
densities using the equivalent values for the S\'ersic $m$ and Einasto
$n$ indices for fixed values of $\Upsilon I_{{\text{e}}}$, $R_{{\text{e}}}$,
$\rho_{0}{r_{{\text{s}}}}$ and ${r_{{\text{s}}}}$,
showing that both profiles have a similar behaviour. However, we noted
that for the Einasto profile the external wings are more spread out
and it seems to be less sensitive than the S\'ersic profile to the
value of the surface mass density for a given Einasto index and radius 
in the inner region. Also, we find that the Einasto profile is more
\textquoteleft cuspy\textquoteright$\;$than the S\'ersic profile:
the former has higher values of the central surface mass density.
These features are of key importance, because it is in this region
that the lensing effect is more important, and these dissimilarities
in the surface mass densities imply a difference in the lensing properties
of both profiles. This result agrees with previous work of \citet{2005MNRAS.358.1325C}
and \citet{2010MNRAS.405..340D}.

Our results are the first step in studying the properties of the Einasto
profile using analytical means. The constant increase of computational
power opens the possibility of using more realistic and sophisticated
profiles like the Einasto profile in cosmological studies, where our
results may apply. For example, they can be used in strong- and
weak-lensing studies of galaxies and clusters, where dark matter is
believed to be the main mass component and the mass distribution can
be assumed to be given by an Einasto profile. The better performance
in cosmological $N$-body simulations of the Einasto profile \citep{2004MNRAS.349.1039N,2006AJ....132.2685M,2008MNRAS.387..536G,2008MNRAS.388....2H,2009MNRAS.398L..21S,2010MNRAS.402...21N}
makes its inclusion in strong and weak lensing studies very promising.
Recently, \citet{2011AJ....142..109C} analysed the rotation curves
(RC) of spiral galaxies from THINGS (The {\small HI} Nearby Galaxy
Survey, \citep{2008AJ....136.2648D}) and found that the Einasto profile
provides a better match to the observed RC than the NFW profile \citep*{1996ApJ...462..563N,1997ApJ...490..493N}
and the cored pseudo-isothermal profile. Also, \citet{2011arXiv1112.3120D}
modelled the surface brightness profiles of a sample of elliptical
galaxies of the Virgo cluster using a multi-component Einasto profile
based on an analytical approximation, and obtained a good fit for shallow-cusp
and steep-cusp galaxies with fit residual errors lower in
comparison to measurement errors in a wide dynamical radial range.
Additionally, in their models of the most massive galaxies, the outer
components are characterised by being in the range $5\lesssim n\lesssim8$,
and a comparable range is obtained from $N$-body simulations for the
Einasto profile. Our exact analytical results for the spatial and
lensing properties of the Einasto may be used to constrain the value
of the Einasto index and determine if the galaxy or cluster studied
is dark matter dominated or not. More studies like \citet{2011AJ....142..109C}
and \citet{2011arXiv1112.3120D} could help to strengthen
the position of the Einasto profile as a new standard model for dark
matter haloes. Also, increasing the use of the Einasto profile in
new cosmological studies could provide progress towards a solution
to the cusp-core problem. 

This paper continues the effort initiated by \citet{2011A&A...525A.136B}
and \citet{2011A&A...534A..69B} to advocate the use of the Fox $H$
and Meijer $G$ functions in theoretical astrophysics, in particular
for studying the analytical properties of density models like the S\'ersic
model, where no additional analytical progress could be made until
the Mellin integral transform formalism was applied. We hope that
our work has again demonstrated the usefulness of the Fox $H$ and
Meijer $G$ functions as tools for analytical work. 

\begin{acknowledgements} ERM and FFA wish to thank H.\ Morales and R.\
  Carboni for critical reading. This research has made use of NASA's
  Astrophysics Data System Bibliographic Services.  GG is a
  postdoctoral researcher of the FWO-Vlaanderen (Belgium). Moreover,
  we would like to thank the referee for valuable suggestions on the
  manuscript.  \end{acknowledgements}

\bibliographystyle{aa}
\addcontentsline{toc}{section}{\refname}\bibliography{my_bib}

\end{document}